\documentclass{article}
\usepackage[T1]{fontenc}
\usepackage[utf8]{inputenc}
\usepackage[]{ismir} 
\usepackage{amsmath,cite,url}
\usepackage{amssymb}
\usepackage{graphicx}
\usepackage{color}
\usepackage{booktabs}
\usepackage{multirow}
\usepackage{svg}
\usepackage{subcaption}
\title{Video-Guided Text-to-Music Generation Using Public Domain Movie Collections}

 \multauthor
   {Haven Kim$^1$ \hspace{0.6cm} Zachary Novack$^1$ \hspace{0.6cm} Weihan Xu$^2$ \hspace{0.6cm} Julian McAuley$^1$ \hspace{0.6cm} Hao-Wen Dong$^3$}
   {$^1$ University of California San Diego\\
   $^2$ Duke University\\
   $^3$ University of Michigan\\
}   

\def\authorname{Haven Kim, Zachary Novack, Weihan Xu, Julian McAuley, and Hao-Wen Dong}

\PassOptionsToPackage{table,xcdraw}{xcolor}
\usepackage{pifont}
\usepackage{enumitem}
\setlist{parsep=0ex,topsep=0.5ex,itemsep=0ex,leftmargin=2em}
\usepackage{inconsolata}
\usepackage{microtype}

\begin{document}

\maketitle

\begin{abstract}

Despite recent advancements in music generation systems, their application in film production remains limited, as they struggle to capture the nuances of real-world filmmaking, where filmmakers consider multiple factors—such as visual content, dialogue, and emotional tone—when selecting or composing music for a scene. This limitation primarily stems from the absence of comprehensive datasets that integrate these elements. To address this gap, we introduce Open Screen Soundtrack Library (OSSL), a dataset consisting of movie clips from public domain films, totaling approximately 36.5 hours, paired with high-quality soundtracks and human-annotated mood information. To demonstrate the effectiveness of our dataset in improving the performance of pre-trained models on film music generation tasks, we introduce a new video adapter that enhances an autoregressive transformer-based text-to-music model by adding video-based conditioning. Our experimental results demonstrate that our proposed approach effectively enhances MusicGen-Medium in terms of both objective measures of distributional and paired fidelity, and subjective compatibility in mood and genre. To facilitate reproducibility and foster future work, we publicly release the dataset, code, and demo~\footnote{Dataset: \url{https://havenpersona.github.io/ossl-v1}\\Code:  \url{https://github.com/havenpersona/ossl-v1}\\Demo:  \url{https://havenpersona.github.io/demo/ismir2025}}.

\end{abstract}

\section{Introduction}
Music plays a crucial role in films, shaping its artistic quality and influencing its commercial success~\cite{millet2021soundtrack}. A well-composed soundtrack enhances the emotional depth of a scene, guiding audience perception and engagement~\cite{thao2019multimodal, won2021emotion, thao2021attendaffectnet, chua2022predicting}.
Despite recent advancements in music generation systems, significant challenges remain in adapting these technologies for film production, as these systems are not designed to align with the real-world practices of film music composition, where multiple elements—such as visual content, dialogue, and emotional tone—are considered altogether~\cite{xu2022analysis}.

A major obstacle is the lack of comprehensive datasets containing movie clips paired with their corresponding soundtracks. While some existing film datasets are derived from commercial movies, many are no longer available for download~\cite{hollywood2, movieqa, movienet}, and some provide only video embeddings rather than raw movie clips~\cite{soldan2022mad}, making it challenging to distinguish segments containing music from those that do not. The remaining datasets do not include isolated soundtrack stems~\cite{rohrbach2015dataset,vicol2018moviegraphs,curtis2020hlvu,bain2020condensed}. While we acknowledge the possibility of constructing a dataset using source-separated music, as done in previous approaches~\cite{xu2024teasergengeneratingteaserslong}, we find the quality of source-separated music to be suboptimal, as it occasionally includes incorrect audio such as voice or sound effects, making it unsuitable for our use case.

Another missing element that plays a pivotal role in making film music is mood information, as dialogues and scenes alone are often insufficient to fully convey emotions, which explains why scripts typically include not only dialogues but also separate mood descriptions. However, we find that this aspect remains largely unaddressed in existing resources, with no publicly available film datasets annotated with mood information. 

To bridge this gap, we introduce the \textbf{Open Screen Soundtrack Library (OSSL)}, a dataset comprising approximately 36.5 hours of movie clips from public domain films, each paired with their corresponding soundtracks and mood annotations. The dataset is created by automatically identifying timestamps of musical segments within the films, mapping these segments to the soundtracks, and then verifying the mappings manually. To demonstrate the effectiveness of this dataset for enhancing music generation for movie clips, we incorporate a video adapter into a text-to-music generation model (MusicGen)~\cite{musicgen} and fine-tune it on our dataset. We evaluate our models on both public domain films and commercial films, which we call OSSL Evaluation Set - Public and Commercial (OES-Pub and OES-Com), respectively, in order to ensure that our approach can handle various kinds of data. Our objective and subjective evaluations demonstrate that we successfully extend a text-to-music generation model, MusicGen-Medium, to handle both textual and video inputs for enhanced film music generation.

For reproducibility and accessibility, we publicly release our main dataset, OSSL, as well as the evaluation sets, OES-Pub and OES-Com.

\begin{figure*}
    \centering
    \includegraphics[width=0.81\textwidth]{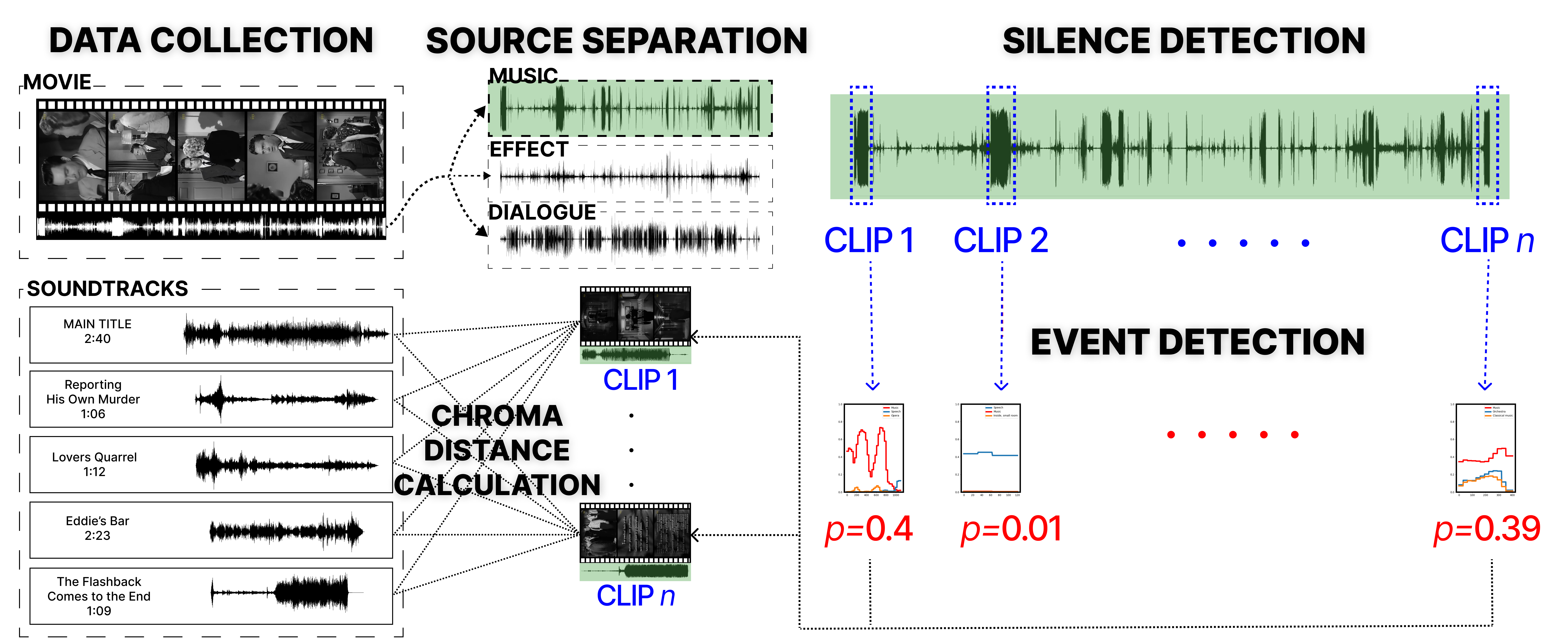}
    \vspace{-2mm}
    \caption{Illustration of our methodology for constructing the OSSL dataset. We collect publicly available movies and their soundtracks. We separate movie audio into music and other elements, and extract music using a silence detection algorithm. An event detection algorithm ensures the extracted audio is truly music. We then calculate chroma distance to match clips with their corresponding soundtracks, assigning the closest match. Human inspection verifies these mappings.}
    \vspace{-2mm}
    \label{fig:dataset-construction}
\end{figure*}

\section{Related Work}

\vspace{0.5ex}
\noindent\textbf{Audio-Domain Music Generation.}\quad  Contemporary music generation architectures in the audio domain predominantly follow two distinct paradigms. The first employs neural codecs to transform digital audio signals into discrete tokens, enabling transformer-based models~\cite{vaswani2017attention} to generate music by learning token distributions from prompts such as text~\cite{musicgen, agostinelli2023musiclm, lan2024musicongen} or video~\cite{tian2024vidmuse, su2024v2meow, zuo2025gvmgen}. The second paradigm leverages diffusion-based frameworks, where diffusion models
are trained to generate audio signals, conditioned on prompts such as textual descriptions~\cite{forsgren2022riffusion, huang2023noise2music, schneider2023mo, lam2023efficient, melechovsky2023mustango, karchkhadze2024multi, evans2024stable} or video frames~\cite{lin2024vmas, muvi2024}. Our models are built upon MusicGen~\cite{musicgen}, a text-to-music generation framework based on the former approach.

\vspace{0.5ex}
\noindent\textbf{Multimodal Music Generation.}\quad Most existing audio-domain music generation models support purely unimodal conditions, such as text~\cite{evans2024stable, agostinelli2023musiclm}, or have limited multimodal support for text along with audio~\cite{musicgen, agostinelli2023musiclm, jasco} or abstract musical signals~\cite{Wu2023MusicControlNet, Novack2024Ditto, Novack2024Ditto2}. In constrast, work on conditioning on \emph{visual} signals is more limited. One example is a transformer-based model conditioned on text, speech, images, and videos, which generates discrete music tokens that are later converted into raw waveforms from embeddings obtained from pre-trained speech, image, and video encoders~\cite{liu2024mumu}. An alternative approach involves converting visual inputs into detailed textual descriptions, and feeding them into a model. This enables the use of diverse modalities, including text, videos, and images as inputs~\cite{wang2024multimodal}. Another recent study trained a video-to-music generation model on a diverse set of video features, while integrating textual conditioning to enable high-level control through textual embeddings derived from a pre-trained text encoder~\cite{su2024v2meow}. On the other hand, our approach focuses on integrating a video adapter into a pre-trained text-to-music generation model in order to build a music generation model conditioned on multimodal inputs.

\vspace{0.5ex}
\noindent\textbf{Video-Conditioned Music Generation.}\quad One of the earliest contributions to video-to-music generation utilized human pose features extracted by pre-trained models, in order to generate plausible music for video clips containing individuals playing musical instruments~\cite{gan2020foley}. In contrast, another early study employed self-defined handcrafted features to capture relevant video attributes~\cite{di2021video}. More recently, the field has shifted toward leveraging embeddings derived from pre-trained video encoders~\cite{zhuo2023video, tian2024vidmuse, su2024v2meow, zuo2025gvmgen}, both in the audio and symbolic domain. This approach has been integrated into our models, enabling them to perform cross-attention on embeddings derived from a pre-trained video encoder designed to capture both spatial and temporal video information~\cite{arnab2021vivit}.


\vspace{0.5ex}
\noindent\textbf{Adapter Mechanisms.}\quad
Adapter mechanisms, which integrate compact and trainable modules into pre-trained models, were originally introduced to enhance transfer learning efficiency in natural language processing (NLP)\cite{houlsby2019parameter}. This approach is commonly employed to adapt pre-trained backbone models for various tasks~\cite{pfeiffer2020adapterfusion, pfeiffer2020adapterhub}.

Our work is particularly inspired by methods that modify pre-trained models to accommodate external conditions with new modalities. Notably, prior research has demonstrated the effectiveness of lightweight adapters in incorporating image prompts into text-to-image diffusion models~\cite{ye2023ip, mou2024t2i}. Building on this idea, subsequent work has extended the approach by using adapters to integrate audio prompts into diffusion-based text-to-music generation models~\cite{tsai2024audio}. On the other hand, our approach attempts to integrate video prompts into autoregressive transformerr-based text-to-music generation models.

\begin{figure}
    \centering
    \includegraphics[width=0.80\columnwidth]{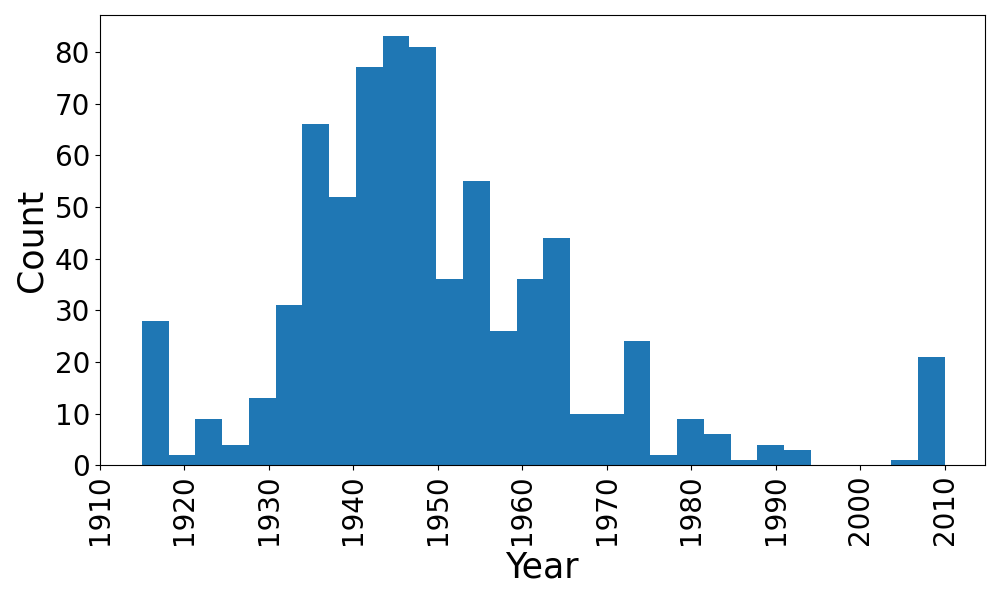}
    \vspace{-2mm}
    \caption{Distribution of release years for the OSSL dataset}
    \vspace{-2mm}
    \label{fig:year}
\end{figure}

\section{Dataset Construction}

\begin{table}[]
\centering
\resizebox{0.99\linewidth}{!}{
\begin{tabular}{@{}ccccccc@{}}
\toprule
Dataset & Audio & MIDI & \begin{tabular}[c]{@{}c@{}}Self-\\ Hosted\end{tabular} & \begin{tabular}[c]{@{}c@{}}Mood\end{tabular} & \begin{tabular}[c]{@{}c@{}}Video\\ Content\end{tabular} & \begin{tabular}[c]{@{}c@{}}Length\\ (Hours)\end{tabular} \\ \midrule
HIMV-200K~\cite{himv2017} & \color[HTML]{009901} \ding{51} & \color[HTML]{FE0000} \ding{55} & \color[HTML]{FE0000} \ding{55} & \color[HTML]{FE0000} \ding{55} & \begin{tabular}[c]{@{}c@{}}Music Video,\\ User-Generated Video\end{tabular} & - \\
URMP~\cite{urmp2019} & \color[HTML]{009901} \ding{51} & \color[HTML]{009901} \ding{51} & \color[HTML]{FE0000} \ding{55} & \color[HTML]{FE0000} \ding{55} & Music Performance & 33.5 \\
TikTok~\cite{tiktok2022} & \color[HTML]{009901} \ding{51} & \color[HTML]{FE0000} \ding{55} & \color[HTML]{FE0000} \ding{55} & \color[HTML]{FE0000} \ding{55} & Dance Video & 1.5 \\
AIST++~\cite{aist2021} & \color[HTML]{009901} \ding{51} & \color[HTML]{FE0000} \ding{55} & \color[HTML]{009901} \ding{51} & \color[HTML]{FE0000} \ding{55} & 3D Dance Motion & 5.2 \\
SymMV~\cite{vtm_symbolic_2023} & \color[HTML]{009901} \ding{51} & \color[HTML]{009901} \ding{51} & \color[HTML]{FE0000} \ding{55} & \color[HTML]{FE0000} \ding{55} & Music Video & 76.5 \\
MuVi-Sync~\cite{Kang_2024} & \color[HTML]{009901} \ding{51} & \color[HTML]{009901} \ding{51} & \color[HTML]{FE0000} \ding{55} & \color[HTML]{FE0000} \ding{55} & Music Video & - \\
BGM909~\cite{bgm909_2024} & \color[HTML]{009901} \ding{51} & \color[HTML]{009901} \ding{51} & \color[HTML]{FE0000} \ding{55} & \color[HTML]{FE0000} \ding{55} & Music Video & - \\
NES-VMDB~\cite{nes_2024} & \color[HTML]{FE0000} \ding{55} & \color[HTML]{009901} \ding{51} & \color[HTML]{FE0000} \ding{55} & \color[HTML]{FE0000} \ding{55} & Gameplay Video & 474.0 \\
OSSL (Ours) & \color[HTML]{009901} \ding{51} & \color[HTML]{FE0000} \ding{55} & \color[HTML]{009901} \ding{51} & \color[HTML]{009901} \ding{51} & Films & 36.5 \\ \bottomrule
\end{tabular}
}
\caption{Comparison of video-music datasets available as of June 2025. The proposed OSSL dataset is the first self-hosted video-music dataset (i.e., without requiring separate download procedures via YouTube URLs or sharing requests) which includes mood annotations.}
\label{tab:comparison}
\vspace{-3mm}
\end{table}

\subsection{Open Screen Soundtrack Library (OSSL)}
We introduce the Open Screen Soundtrack Library (OSSL), a collection of movie clips with their corresponding soundtracks and associated metadata, including mood annotations. We provide an overview of the comparison of video-music datasets in Table~\ref{tab:comparison} and an illustration of our dataset construction methodology in Figure~\ref{fig:dataset-construction}.

\vspace{0.5ex}
\noindent\textbf{Data Collection.}\quad We compile a list of public domain films along with metadata (e.g., title, release date, and genres) and obtain complete versions from YouTube. To achieve the highest music quality by obtaining soundtrack stems without unnecessary noise, we download soundtracks, instead of source-separated music, for each film from YouTube, guided by IMDB\footnote{An online database on films. \url{https://www.imdb.com}} metadata. The frame rate and resolution for each video are 25fps and 960x720, respectively, and the sampling rate for each soundtrack is 44.1kHz.

\vspace{0.5ex}
\noindent\textbf{Musical Segment Identification.}\quad 
To identify the timestamps of segments with soundtracks within the films, we employ a pre-trained source separation model~\cite{solovyev2023benchmarks} trained to decompose movie audio into three components: music, effect, and dialogue. After source separation, we apply a silence detection algorithm to music parts using \texttt{pyAudioAnalysis}~\cite{giannakopoulos2015pyaudioanalysis}, with a frame length and step size of 20ms and a threshold scaling factor of 0.2. We define a segment as a clip if it is longer than 10 seconds and contains no silence longer than 1 second.

Following this, we verify the presence of musical content using an event detection algorithm~\cite{kong2020panns}, as the extracted soundtrack may contain non-musical elements owing to the limitations of current source separation models. Specifically, we apply the event detection algorithm to each clip and retain only those where the average probability of containing a musical event exceeds 0.3. (This threshold value is determined empirically by testing different values across multiple samples.)

\vspace{0.5ex}
\noindent\textbf{Movie Clips-Soundtracks Mapping.}\quad  Because a movie typically contains multiple soundtracks, it was essential to determine which soundtrack each movie clip corresponds to. The most effective method we found is using chroma similarity~\footnote{We attempted a fingerprinting approach (https://github.com/worldveil/dejavu) to test mapping 20 clips from the movie ``D.O.A'' against 24 soundtracks. However, this method produced inaccurate mappings in most instances, achieving only one correct identification out of 20 samples—a failure rate of 95\%. Our chroma similarity approach, on the other hand, accurately mapped 17 clips to their corresponding soundtracks, yielding a success rate of 85\%. }. Specifically, we compare the chroma features of source-separated soundtracks from movie clips with those of each soundtrack in the film, assigning clips to the soundtrack with the minimum chroma distance. 

\vspace{0.5ex}
\noindent\textbf{Manual Quality Inspection.}\quad The resulting movie clips are manually assessed by human evaluators, and any clips with incorrect mappings (i.e., when a clip is paired to a wrong soundtrack) are excluded from our dataset.

\begin{table}[]
\centering
\resizebox{0.9\linewidth}{!}{
\begin{tabular}{@{}cccc@{}}
\toprule
\textbf{Attributes} & \textbf{OSSL} & \textbf{OES-Pub} & \textbf{OES-Com} \\ \midrule
Number of samples & 736 & 100 & 100\\
Number of unique films & 299 & 76 & 37\\
Average clip duration & 178.47 sec & 30 sec & 30 sec\\
Total duration & 36.49 hours & 0.83 hours & 0.83 hours\\ \bottomrule
\end{tabular}
}
\caption{Statistical overview of OSSL, OES-Pub, and OES-Com}
\vspace{-5mm}
\label{tab:dataset_statistics}
\end{table}

\vspace{0.5ex}
\noindent\textbf{Mood Annotation.}\quad The final stage of our dataset construction involves annotating mood information for each movie clip. We classify mood into four categories based on a previously suggested taxonomy—Russell's 4Q, where the four classes are one of the HVHA (high valence, high arousal), HVLA (high valence, low arousal, LVHA (low valence, high arousal), and LVLA (low valence, low arousal)~\cite{russell1980circumplex, hung2021emopia}.

We employ two human annotators. We first provide them with a brief explanation of the concepts of valence and arousal in the context of music and then ask them to independently annotate the movie clips using the four mood categories. In the majority of cases (89.9\% of samples), both annotators assign the same label to a movie clip. When agreement occurs, we retain the assigned annotation. If they disagree, they discuss their choices until they reach a consensus. As a result, we obtain 276 movie clips classified as happy, 30 as sad, 315 as nervous, and 115 as peaceful.

\subsection{OSSL Evaluation Set (OES)}

We evaluate our models on two distinct datasets: OSSL Evaluation Set-Public (OES-Pub), comprising of movie clips from public domain films that are not included in OSSL, and OSSL Evaluation Set-Commercial (OES-Com), consisting of commercial films. The evaluation on OES-Pub serves to confirm that our models have effectively learned to generalize from this data distribution. On the other hand, the evaluation on OES-Com shows whether these models can generalize to contemporary commercial films. Notably, OES-Com incorporates a large proportion (89\%) of clips from films released within the past year. Finally, we annotate each movie clip with mood information in the same way as we did for constructing OSSL (\texttt{[MOOD]}).

Finally, we present detailed statistical information for OSSL, OES-Pub, and OES-Com in Table~\ref{tab:dataset_statistics}.

\begin{figure}
    \centering
\includegraphics[width=0.80\linewidth]{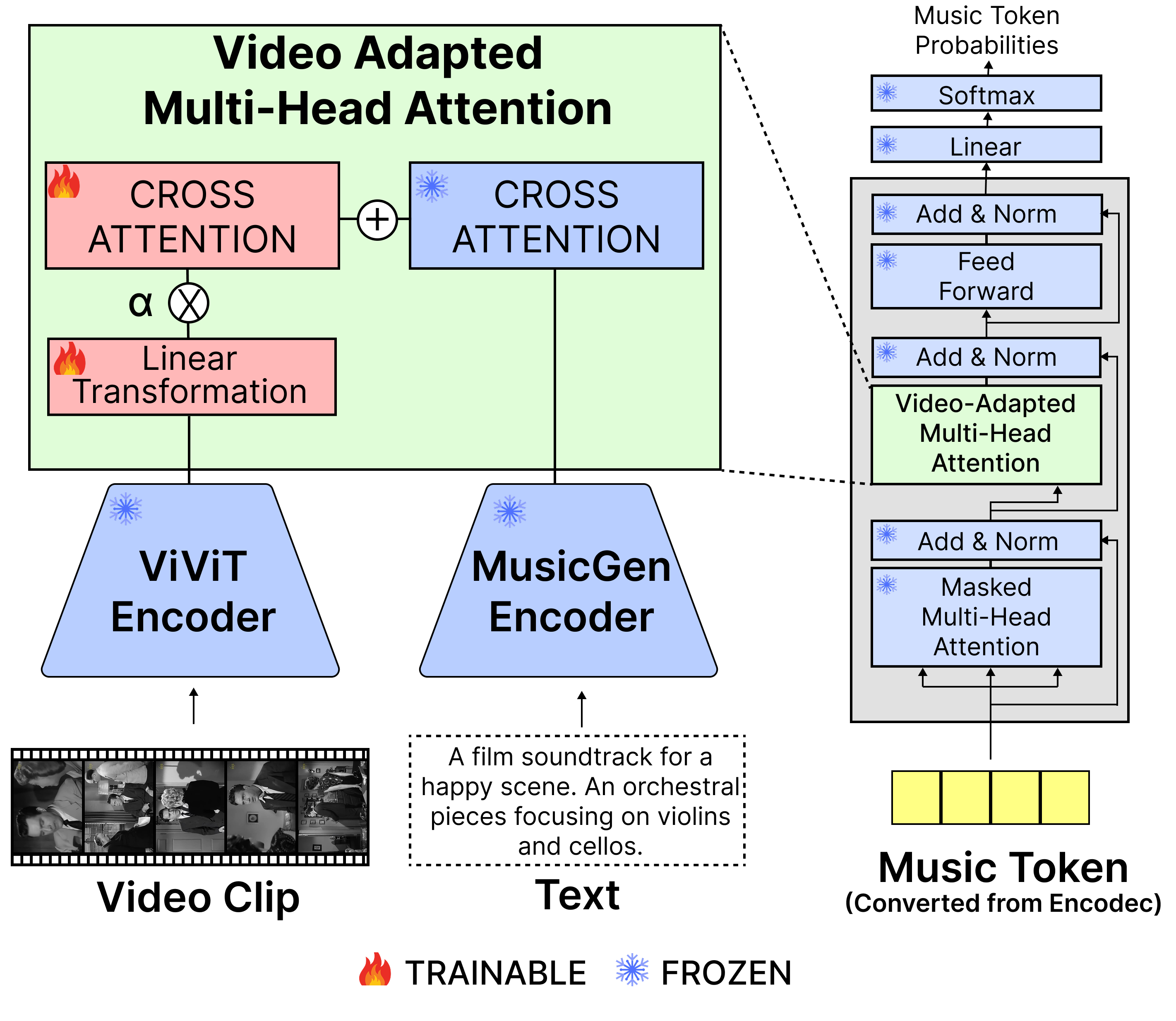}
    \vspace{-2mm}
    \caption{Extending MusicGen: We introduce a video adapter, which applies cross-attention to linearly transformed video embeddings, scaled by $\alpha$, and integrates it into the original cross-attention mechanism.
    A fire icon denotes trainable components, while a snowflake icon indicates frozen components.}
    \label{fig:model_architecture}
    \vspace{-2mm}
\end{figure}

\section{Model Architecture}

In this section, we present our methodology for integrating a video adapter into an existing text-to-music generation model, MusicGen~\cite{musicgen}, along with its illustration in Figure~\ref{fig:model_architecture}.

MusicGen~\cite{musicgen} is an
autoregressive
transformer~\cite{vaswani2017attention}-based model that generates discrete tokens which are subsequently converted into audio signals by a neural codec~\cite{defossez2022high} from textual prompts. Each attention head in the original model's cross attention modules is initially defined as:
\begin{equation}
\text{head}_i = \text{Attention}(x\mathbf{W_i^{(q)}}, z^t\mathbf{W_i^{(k)}}, z^t\mathbf{W_i^{(v)}})
\end{equation}
In this formulation, $x$ represents the decoder's current hidden states, $z^t$ means the text encoder's output, and $i$ is the index of an attention head.

To implement our video adapter, we leverage a pretrained video encoder. Specifically, we choose the ViViT\footnote{\texttt{google/vivit-b-16x2-kinetics400}}~\cite{arnab2021vivit}
, a transformer-based video understanding model pretrained on a large-scale dataset~\cite{kay2017kineticshumanactionvideo}, which provides a strong foundation for our task of film video understanding. We first obtain the video embeddings ($z_v\in \mathbb{R}^n$) using it. Subsequently, we apply an affine linear transformation $\mathbf{X}\in \mathbb{R}^{m\times n}$ to adjust the dimension of video embeddings from the model's original dimension $n$ to the dimension that is compatible with our text embeddings $m$ ($\tilde{z_v} = \mathbf{X} * z_v$).

We modify the cross-attention layer, which originally processes single modalities, to incorporate video embeddings, thereby enabling the model to attend to multimodal contexts. Specifically, we augment the original cross-attention mechanism with a video-conditioned component, where each attention head computes its output by leveraging both text and video modalities.
\vspace{-2mm}
\begin{equation}
\begin{aligned}
\text{head}_i = \text{Attention}(x\mathbf{W_i^{(q)}}, z_t\mathbf{W_i^{(k)}}, z_t\mathbf{W_i^{(v)}}) \\
+ \alpha \times \text{Attention}(x\mathbf{\Tilde{W_i}^{(q)}}, \Tilde{z_v}\mathbf{\Tilde{W_i}^{(k)}}, \Tilde{z_v}\mathbf{\Tilde{W_i}^{(v)}})
\end{aligned}
\end{equation}
\vspace{-3mm}

where $x$ represents the decoder's current hidden states, $\Tilde{z_v}$ represents the video embeddings with adjusted dimensions, and $\alpha$ is a trainable parameter. 

During training, only the newly introduced parameters \(\mathbf{\Tilde{W_i}^{(q)}}\), \(\mathbf{\Tilde{W_i}^{(k)}}\), \(\mathbf{\Tilde{W_i}^{(v)}}\), \(\alpha\), and \(\mathbf{X}\) are optimized after random initialization, while all other components of the model remain frozen.

\section{Experimental Setting}

\subsection{Comparison Models}

We fine-tune MusicGen-Small and MusicGen-Medium with video adapters,\footnote{Due to computational constraints, we did not conduct experiments on MusicGen-Large.} as described in the previous section. Here, \textsc{S-Multi} and \textsc{M-Multi} denote these models, where ``S'' and ``M'' stand for ``Small'' and ``Medium,'' respectively.

As baselines, we use the original MusicGen-Small and MusicGen-Medium models, which generate results solely based on text prompts. We denote these models as \textsc{S-Base} and \textsc{M-Base}, respectively.

To assess the effectiveness of video adapters, we also compare them against models fine-tuned on our dataset using only textual prompts (i.e., without video adapters). These models are denoted as \textsc{S-Text} and \textsc{M-Text}. Unlike \textsc{S-Multi} and \textsc{M-Multi}, where existing parameters are frozen and new parameters are trained, \textsc{S-Text} and \textsc{M-Text} do not incorporate new parameters. Instead, we apply Low-Rank Adaptation~\cite{hu2021lora} when fine-tuning these models.

\subsection{Dataset Preprocessing}

Since the default maximum length of MusicGen is 30 seconds, we train our models to generate only the first 30 seconds of music for the first 30 seconds of each video clip. This is because we detected the start time of each segment, so it prevents generating from the middle of a track. Because MusicGen produces raw audio at a sampling rate of 32,000 Hz, all soundtracks are resampled to this rate before training. Additionally, we normalize all soundtracks so that their maximum amplitude is always one.

\subsection{Text Prompts Design}

Prompts have been suggested to be a significant factor influencing the outcomes of generative models~\cite{brown2020language, liu2022design, white2023prompt}. In our experiments, text prompts serve not only as primary information for fine-tuning text-based models (\textsc{S-Text} and \textsc{M-Text}) but also as a basis for inference in baseline models (\textsc{S-Base} and \textsc{M-Base}). To ensure a fair comparison, therefore, we carefully design them to fully leverage the capabilities of text-to-music generation models,.

We experiment with three types of textual information: (1) mood annotations from our dataset, (2) movie genre labels (e.g., thriller, romance), and (3) LLM-generated music descriptions obtained using an open-source music captioning model~\cite{doh2023lp}. We observe that while both mood annotations and LLM-generated music captions subjectively improve generation quality, adding genre information often has a negative impact. For instance, when using the prompt, ``A film soundtrack for a peaceful scene in a thriller movie,'' our baseline models, \textsc{S-Base} and \textsc{M-Base}, overemphasizes the keyword ``thriller,'' producing music unsuitable for peaceful scenes. As a result, our final structured template includes only mood information and LLM-generated music captions:
``A film soundtrack for a \texttt{[MOOD]} scene. \texttt{[CAPTION]}.'' Here, \texttt{[MOOD]} is a natural language description of our Russell's Q4-based mood annotation, which is one of happy (high valence, high arousal), sad (low valence, low arousal), nervous (low valence, high arousal), or peaceful (high valence, low arousal). To generate music captions (\texttt{[CAPTION]}), we extract the first 30 seconds of each soundtrack in our dataset, divide it into three 10-second segments, and generate a caption for each using a music captioning model~\cite{doh2023lp}, as we recognize that film music evolves over time. We then use a commercial LLM, specifically Claude 3.5 Sonnet, to summarize these captions with the prompt:
``Summarize the description of each song in one sentence from 0 to 30 seconds.'' Consequently, the resulting captions concisely describe how the music changes over time, providing a brief musical summary (e.g., `The piece transitions from cembalo to marimba, concluding with a Tibetan singing bowl and animal sounds').

When designing text prompts for OES-Pub and OES-Com, we use source-separated music to generate \texttt{[CAPTION]}, unlike OSSL, which features original soundtrack stems. This results in the music captioning model frequently describing the audio as having poor recording quality. To mitigate this, we explicitly instruct Claude 3.5 Sonnet to exclude any mention of audio quality when summarizing the captions for OES-Pub and OES-Com.

\subsection{Training Details}
Our models are trained using the AdamW optimizer~\cite{adamw} \((\beta_1=0.9, \beta_2=0.999)\) with a weight decay of \(1 \times 10^{-2}\). The initial learning rate is set to \(1 \times 10^{-4}\) and scheduled using a cosine annealing strategy~\cite{loshchilov2016sgdr} with a linear warm-up phase. The OSSL dataset is split into training and validation sets in a 9:1 ratio. To prevent overfitting, we stop training a model when the model does not improve for three epochs during the validation stage. Due to computational constraints, the batch size is set to 1. Training is conducted on a single NVIDIA A6000 GPU.

\subsection{Objective Evaluation Metrics}


\vspace{0.5ex}
\noindent\textbf{Distributional Fidelity.}\quad As common in most audio-domain TTM research~\cite{Novack2025Presto, nistal2024diff}, we assess the quality of our outputs at the distributional level by comparing generations from our model against a high-quality reference set, here comprised of 5K commercial soundtracks. To do so, we first extract embedding from our generated audio for each method and the reference set using CLAP~\cite{wu2023large}. With these embeddings, we then compute Frechet Audio Distance (FAD)~\cite{kilgour2018fr} and Precision~\cite{ferjad2020icml} for this purpose. FAD compares distributional distance by fitting a high dimensional Gaussian to each dataset and measuring the Frechét distance between them, while Precision uses a k-NN estimate of the reference set's distribution and measures how many generated samples lie in the estimated manifold.

\vspace{0.5ex}
\noindent\textbf{Paired Fidelity.}\quad As our evaluation sets, OES-Pub and OES-Com, also contain paired video-music data in the form of source-separated musical tracks from each movie clip, we are also able to directly assess how well our models recreate the reference music on a paired sample-by-sample basis. Specifically, here we measure the CLAP Audio Similarity and Kullback-Leibler (KL) Divergence between the generated music and the reference music for this purpose. The CLAP Audio Similarity is calculated as a cosine similarity between the CLAP embeddings of the generated and reference samples for each video clip. The KL divergence is calculated using the estimated distributions the generated and reference samples with the PaSST audio classifier~\cite{koutini2021efficient}.

\vspace{0.5ex}
\noindent\textbf{Sample Diversity.}\quad To evaluate the diversity of the generated samples, we employ Recall~\cite{ferjad2020icml}, following prior work in music generation
\cite{nistal2024diff, Novack2025Presto}. Using the same embedding model (CLAP) and reference/generated datasets from Distributional Fidelity, we use a k-NN estimate of each \emph{generated} distribution and calculate the fraction of real samples that lie in the generated manifold.

\begin{table*}[]
\centering
\resizebox{1.0\textwidth}{!}{
\begin{tabular}{@{}lcclclclclclcccc@{}}
\toprule
\multirow{4}{*}{} & \multirow{4}{*}{\begin{tabular}[c]{@{}c@{}}OSSL\\ Fine-\\ tuned\end{tabular}} & \multirow{4}{*}{\begin{tabular}[c]{@{}c@{}}Video\\ Adapter\\ Inte-\\grated\end{tabular}} & \multicolumn{10}{c}{Objective} & \multicolumn{3}{c}{Subjective} \\ \cmidrule(l){4-16} 
 &  &  & \multicolumn{4}{c}{\begin{tabular}[c]{@{}c@{}}Distributional Fidelity\end{tabular}} & \multicolumn{4}{c}{\begin{tabular}[c]{@{}c@{}}Paired Fidelity\end{tabular}} & \multicolumn{2}{c}{\begin{tabular}[c]{@{}c@{}}Diversity\end{tabular}} & \multicolumn{3}{c}{\begin{tabular}[c]{@{}c@{}}Human Ratings\end{tabular}}  \\ \cmidrule(l){4-16} 
 &  &  & \multicolumn{2}{c}{FAD ↓} & \multicolumn{2}{c}{Precision ↑} & \multicolumn{2}{c}{Similarity ↑} & \multicolumn{2}{c}{KL ↓} & \multicolumn{2}{c}{Recall ↑} &  {Mood ↑} & {Genre ↑}& {Quality ↑} \\ \cmidrule(l){4-16} 
 &  &  & \multicolumn{1}{c}{pub} & com & \multicolumn{1}{c}{pub} & com & \multicolumn{1}{c}{pub} & com & \multicolumn{1}{c}{pub} & com & \multicolumn{1}{c}{pub} & com &  avg $\pm$ CI&  avg $\pm$ CI &  avg $\pm$ CI \\ \midrule
\textsc{S-Base} & {\color[HTML]{FE0000} \ding{55}} & {\color[HTML]{FE0000} \ding{55}} &  64.91 & 77.99 &  22.00 & 14.00 &  41.55 & 34.77 &  1.20 & 1.93 &  4.78 & 6.20 & 4.53 $\pm$ 0.91 & 5.27 $\pm$ 1.04 & 5.67 $\pm$ 0.99 \\
\textsc{S-Text} & {\color[HTML]{009901} \ding{51}} & {\color[HTML]{FE0000} \ding{55}} & 61.98 & 75.75 &  20.00 & 14.00 &  42.44 & 34.81 &  1.13 & 1.97 & 9.90 & \textbf{19.00} & 5.77 $\pm$ 1.73 & 6.00 $\pm$ 1.11 & 6.20 $\pm$ 0.76 \\
\textsc{S-Multi} & {\color[HTML]{009901} \ding{51}} & {\color[HTML]{009901} \ding{51}} &  64.39 & 75.59 &  16.00 & 14.00 &  43.36 & 36.09 &  1.15 & 1.98 & 7.50 & 3.30 & 4.93 $\pm$ 0.99 & 5.87 $\pm$ 1.00 & 6.10 $\pm$ 0.91 \\
\textsc{M-Base} & {\color[HTML]{FE0000} \ding{55}} & {\color[HTML]{FE0000} \ding{55}} &  60.91 & 76.79 &  21.00 & 12.00 &  43.61 & 34.39 &  1.06 & 1.88 &  8.04 & 8.46 & 5.13 $\pm$ 0.94 & 5.60 $\pm$ 1.12 & 6.20 $\pm$ 0.84 \\
\textsc{M-Text} & {\color[HTML]{009901} \ding{51}} & {\color[HTML]{FE0000} \ding{55}} &  61.15 & 77.79 &  24.00 & 17.00 & \textbf{45.31} & 33.72 &  1.04 & 1.90 &  7.28 & 13.78 & 5.20 $\pm$ 0.99 & 6.03 $\pm$ 0.98 & 6.00 $\pm$ 1.02 \\
\textsc{M-Multi} & {\color[HTML]{009901} \ding{51}} & {\color[HTML]{009901} \ding{51}} & \textbf{59.51} & \textbf{73.26} & \textbf{25.00} & \textbf{21.00} & \textbf{45.31} & \textbf{36.25} & \textbf{1.00} & \textbf{1.81} & \textbf{9.96} & 8.72 & 6.20 $\pm$ 1.05 & 6.70 $\pm$ 1.06 & 7.07 $\pm$ 0.93 \\ \bottomrule
\end{tabular}
}
\vspace{-3mm}
\caption{Evaluation results. Objective metrics include FAD, Precision, CLAP Audio Similarity (Similarity), KL Divergence, and Recall. Subjective metrics include human ratings for mood, genre, and audio quality. \textit{pub} and \textit{com} indicate results on OES-Pub and OES-Com, respectively; \textit{avg $\pm$ CI} refers to average values and 95\% confidence intervals.}
\vspace{-5mm}
\label{tab:results}
\end{table*}

\subsection{Subjective Survey}

To subjectively evaluate our models, we first select 10 representative samples from our evaluation set, OES-Com. Specifically, the OES-Com is divided into 10 clusters using k-means clustering, based on the CLAP embeddings of source-separated music, and we select the sample closest to the centroid of each cluster. Using these 10 samples, we design a survey in the form of a website and distribute it within our social network, recruiting 15 participants.

The survey procedure is structured as follows: Each participant is randomly assigned 2 out of the 10 samples. First, participants are required to watch the original version of the first movie clip. This step serves two purposes: to familiarize them with the atmosphere of the clip and to standardize the experience for participants regardless of prior exposure to the movie. After viewing the original clip, participants rate the inference results of each model (\textsc{S-Base}, \textsc{S-Text}, \textsc{S-Multi}, \textsc{M-Base}, \textsc{M-Text}, and \textsc{M-Multi}) for the corresponding clip, with the presentation order of the models randomized by the website. Subsequently, participants repeat the process for the second assigned movie clip, watching its original version and rating the inference results of each model, with the randomized order.

The evaluation assesses each model across three dimensions—genre, mood, and audio quality—using a 10-point Likert scale. For genre, participants rate how cinematic the AI-generated music sounds (1: not cinematic at all, 10: very cinematic). For mood, participants evaluate the compatibility of the music with the emotional tone of the movie clip (1: not compatible at all, 10: very compatible). For audio quality, participants provide a subjective assessment of the audio quality (1: very poor quality, 10: very high quality). 

In the following section, we present the average rating values and 95\% confidence intervals for each model, based on each of the three criteria.

\section{Results and Analysis}

We present our comprehensive evaluation results in Table~\ref{tab:results}. 

\vspace{0.5ex}
\noindent\textbf{Distributional Fidelity}\quad Our evaluation on OES-Com reveals that fine-tuning on OSSL enhances distributional fidelity. Specifically, \textsc{S-Text} and \textsc{S-Multi} achieve lower FAD scores compared to \textsc{S-Base} while \textsc{M-Text} and \textsc{M-Multi} exhibit higher Precision scores relative to \textsc{M-Base}, indicating closer alignment with the target distribution. In contrast, the results on OES-Pub show no significant improvement in distributional fidelity. Although \textsc{M-Text} and \textsc{M-Multi} demonstrate increased Precision scores compared to \textsc{M-Base}, \textsc{S-Text} and \textsc{S-Multi} actually experience a decrease in Precision when evaluated on OES-Pub. This discrepancy is expected, as distributional fidelity is calculated using reference embeddings from commercial movie soundtracks. Encouragingly, the ability to improve performance on commercial soundtracks by training on public domain data suggests effective feature transfer across domains. Particularly, \textsc{M-Multi} achieved the lowest FAD scores and highest Precision scroes on both evaluation sets, demonstrating its outstanding performance in distributional fidelity.

\vspace{0.5ex}
\noindent\textbf{Paired Fidelity}\quad Fine-tuning on OSSL consistently increases the CLAP audio similarity between reference and generated music across both OES-Pub and OES-Com datasets. Additionally, training reduces KL divergence when evaluated on OES-Pub, indicating improved alignment in classifier predictions. However, this improvement is less pronounced on OES-Com, suggesting some limitations in generalizing to commercial soundtracks. Notably, \textsc{M-Multi} stands out by achieving significantly lower KL scores on both datasets, highlighting its superior performance in paired fidelity.

\vspace{0.5ex}
\noindent\textbf{Sample Diversity}\quad Training on OSSL generally results in a slight increase in sample diversity
relative to the base models, 
as indicated by Recall scores when evaluated on OES-Pub, with the exception of \textsc{M-Text}, which shows a minor decrease. We observe the significantly higher Recall scores for \textsc{S-Text} and \textsc{M-Text} on OES-Com, despite the use of commercial soundtracks as reference tracks. This finding counters initial concerns about overfitting to patterns unique to public domain data, demonstrating that fine-tuning can enhance diversity even on out-of-domain references. However, \textsc{S-Multi} exhibits a notable drop in diversity, likely due to the added complexity of learning from both text and video inputs. In contrast, the larger \textsc{M-Multi} model avoids this issue and even shows slight increases in Recall scores on both datasets
relative to the baseline,
suggesting that greater model capacity helps mitigate the challenges of multimodal training.

\vspace{0.5ex}
\noindent\textbf{Human Ratings}\quad  Human ratings for mood, genre, and quality exhibit wide 95\% confidence intervals, reflecting considerable variability in subjective assessments. Despite this, models fine-tuned on OSSL generally receive higher average ratings for mood and genre fidelity compared to their baseline counterparts, indicating better alignment with human expectations in these dimensions. However, training has minimal impact on perceptual quality. We observe an interesting trend video adapters: integrating them into smaller models (\textsc{S-Multi}) slightly lowers average ratings across all metrics, whereas in medium-sized models (\textsc{M-Multi}), they enhance performance in mood, genre, and quality. This indicates that the benefits of video integration may be contingent on sufficient model capacity.

\section{Conclusion and Future Work}

In this paper, we introduced the Open Screen Soundtrack Library (OSSL), a dataset comprising movie clips, corresponding soundtracks, and mood annotations. 
To show the effectiveness of our dataset, we adapted a text-to-music generation model with video conditions and fine-tuned it on our dataset. We conducted evaluations both on public domain and commercial films, and the results show the effectiveness of our dataset and architecture, when applied to medium-sized models. However, due to the limited number of participants, the subjective evaluation requires further validation. Despite this, we believe that the OSSL will advance film music research within ethical boundaries and our experiments on the proposed methods show insightful observations for integrating video modalities into existing text-to-music generation models.

\newpage

\section{Ethics Statement}
Our research adheres to ethical principles by ensuring that the construction of Open Screen Soundtrack Library (OSSL) and training methodologies are based solely on copyright-free materials. By publicly releasing our dataset, we aim to promote ethical research practices and encourage the broader community to utilize copyright-free data for model training, fostering transparency and responsible AI development.  

For evaluation, we used commercial movie clips strictly for scientific purposes, with no intent to infringe upon the rights of content producers. We believe our use falls within ethical and fair-use guidelines. However, to respect copyright laws, we will not release video clips of the OES-Com, and instead provide YouTube URLs for video clips for reproducibility.

\bibliography{ISMIRtemplate}

\begin{thebibliography}{10}
\providecommand{\url}[1]{#1}
\csname url@samestyle\endcsname
\providecommand{\newblock}{\relax}
\providecommand{\bibinfo}[2]{#2}
\providecommand{\BIBentrySTDinterwordspacing}{\spaceskip=0pt\relax}
\providecommand{\BIBentryALTinterwordstretchfactor}{4}
\providecommand{\BIBentryALTinterwordspacing}{\spaceskip=\fontdimen2\font plus
\BIBentryALTinterwordstretchfactor\fontdimen3\font minus \fontdimen4\font\relax}
\providecommand{\BIBforeignlanguage}[2]{{%
\expandafter\ifx\csname l@#1\endcsname\relax
\typeout{** WARNING: IEEEtran.bst: No hyphenation pattern has been}%
\typeout{** loaded for the language `#1'. Using the pattern for}%
\typeout{** the default language instead.}%
\else
\language=\csname l@#1\endcsname
\fi
#2}}
\providecommand{\BIBdecl}{\relax}
\BIBdecl

\bibitem{millet2021soundtrack}
B.~Millet, J.~Chattah, and S.~Ahn, ``Soundtrack design: The impact of music on visual attention and affective responses,'' \emph{Applied ergonomics}, vol.~93, p. 103301, 2021.

\bibitem{thao2019multimodal}
H.~T.~P. Thao, D.~Herremans, and G.~Roig, ``Multimodal deep models for predicting affective responses evoked by movies.'' in \emph{ICCV Workshops}, 2019, pp. 1618--1627.

\bibitem{won2021emotion}
M.~Won, J.~Salamon, N.~J. Bryan, G.~J. Mysore, and X.~Serra, ``Emotion embedding spaces for matching music to stories,'' \emph{arXiv preprint arXiv:2111.13468}, 2021.

\bibitem{thao2021attendaffectnet}
H.~T.~P. Thao, B.~Balamurali, G.~Roig, and D.~Herremans, ``Attendaffectnet--emotion prediction of movie viewers using multimodal fusion with self-attention,'' \emph{Sensors}, vol.~21, no.~24, p. 8356, 2021.

\bibitem{chua2022predicting}
P.~Chua, D.~Makris, D.~Herremans, G.~Roig, and K.~Agres, ``Predicting emotion from music videos: exploring the relative contribution of visual and auditory information to affective responses,'' \emph{arXiv preprint arXiv:2202.10453}, 2022.

\bibitem{xu2022analysis}
K.~Xu, ``Analysis of the roles of film soundtracks in films,'' in \emph{2022 International Conference on Comprehensive Art and Cultural Communication (CACC 2022)}.\hskip 1em plus 0.5em minus 0.4em\relax Atlantis Press, 2022, pp. 351--355.

\bibitem{hollywood2}
M.~Marszalek, I.~Laptev, and C.~Schmid, ``Actions in context,'' in \emph{2009 IEEE Conference on Computer Vision and Pattern Recognition}.\hskip 1em plus 0.5em minus 0.4em\relax IEEE, 2009, pp. 2929--2936.

\bibitem{movieqa}
M.~Tapaswi, Y.~Zhu, R.~Stiefelhagen, A.~Torralba, R.~Urtasun, and S.~Fidler, ``Movieqa: Understanding stories in movies through question-answering,'' in \emph{Proceedings of the IEEE conference on computer vision and pattern recognition}, 2016, pp. 4631--4640.

\bibitem{movienet}
Q.~Huang, Y.~Xiong, A.~Rao, J.~Wang, and D.~Lin, ``Movienet: A holistic dataset for movie understanding,'' in \emph{Computer Vision--ECCV 2020: 16th European Conference, Glasgow, UK, August 23--28, 2020, Proceedings, Part IV 16}.\hskip 1em plus 0.5em minus 0.4em\relax Springer, 2020, pp. 709--727.

\bibitem{soldan2022mad}
M.~Soldan, A.~Pardo, J.~L. Alc{\'a}zar, F.~Caba, C.~Zhao, S.~Giancola, and B.~Ghanem, ``Mad: A scalable dataset for language grounding in videos from movie audio descriptions,'' in \emph{Proceedings of the IEEE/CVF Conference on Computer Vision and Pattern Recognition}, 2022, pp. 5026--5035.

\bibitem{rohrbach2015dataset}
A.~Rohrbach, M.~Rohrbach, N.~Tandon, and B.~Schiele, ``A dataset for movie description,'' in \emph{Proceedings of the IEEE conference on computer vision and pattern recognition}, 2015, pp. 3202--3212.

\bibitem{vicol2018moviegraphs}
P.~Vicol, M.~Tapaswi, L.~Castrejon, and S.~Fidler, ``Moviegraphs: Towards understanding human-centric situations from videos,'' in \emph{Proceedings of the IEEE conference on computer vision and pattern recognition}, 2018, pp. 8581--8590.

\bibitem{curtis2020hlvu}
K.~Curtis, G.~Awad, S.~Rajput, and I.~Soboroff, ``Hlvu: A new challenge to test deep understanding of movies the way humans do,'' in \emph{Proceedings of the 2020 International Conference on Multimedia Retrieval}, 2020, pp. 355--361.

\bibitem{bain2020condensed}
M.~Bain, A.~Nagrani, A.~Brown, and A.~Zisserman, ``Condensed movies: Story based retrieval with contextual embeddings,'' in \emph{Proceedings of the Asian Conference on Computer Vision}, 2020.

\bibitem{xu2024teasergengeneratingteaserslong}
\BIBentryALTinterwordspacing
W.~Xu, P.~P. Liang, H.~Kim, J.~McAuley, T.~Berg-Kirkpatrick, and H.-W. Dong, ``Teasergen: Generating teasers for long documentaries,'' 2024. [Online]. Available: \url{https://arxiv.org/abs/2410.05586}
\BIBentrySTDinterwordspacing

\bibitem{musicgen}
J.~Copet, F.~Kreuk, I.~Gat, T.~Remez, D.~Kant, G.~Synnaeve, Y.~Adi, and A.~D{\'e}fossez, ``Simple and controllable music generation,'' \emph{Advances in Neural Information Processing Systems}, vol.~36, 2024.

\bibitem{vaswani2017attention}
A.~Vaswani, ``Attention is all you need,'' \emph{Advances in Neural Information Processing Systems}, 2017.

\bibitem{agostinelli2023musiclm}
A.~Agostinelli, T.~I. Denk, Z.~Borsos, J.~Engel, M.~Verzetti, A.~Caillon, Q.~Huang, A.~Jansen, A.~Roberts, M.~Tagliasacchi \emph{et~al.}, ``Musiclm: Generating music from text,'' \emph{arXiv preprint arXiv:2301.11325}, 2023.

\bibitem{lan2024musicongen}
Y.-H. Lan, W.-Y. Hsiao, H.-C. Cheng, and Y.-H. Yang, ``Musicongen: Rhythm and chord control for transformer-based text-to-music generation,'' \emph{arXiv preprint arXiv:2407.15060}, 2024.

\bibitem{tian2024vidmuse}
Z.~Tian, Z.~Liu, R.~Yuan, J.~Pan, Q.~Liu, X.~Tan, Q.~Chen, W.~Xue, and Y.~Guo, ``Vidmuse: A simple video-to-music generation framework with long-short-term modeling,'' \emph{arXiv preprint arXiv:2406.04321}, 2024.

\bibitem{su2024v2meow}
K.~Su, J.~Y. Li, Q.~Huang, D.~Kuzmin, J.~Lee, C.~Donahue, F.~Sha, A.~Jansen, Y.~Wang, M.~Verzetti \emph{et~al.}, ``V2meow: Meowing to the visual beat via video-to-music generation,'' in \emph{Proceedings of the AAAI Conference on Artificial Intelligence}, vol.~38, no.~5, 2024, pp. 4952--4960.

\bibitem{zuo2025gvmgen}
H.~Zuo, W.~You, J.~Wu, S.~Ren, P.~Chen, M.~Zhou, Y.~Lu, and L.~Sun, ``Gvmgen: A general video-to-music generation model with hierarchical attentions,'' \emph{arXiv preprint arXiv:2501.09972}, 2025.

\bibitem{forsgren2022riffusion}
S.~Forsgren and H.~Martiros, ``Riffusion-stable diffusion for real-time music generation,'' \emph{URL https://riffusion. com}, 2022.

\bibitem{huang2023noise2music}
Q.~Huang, D.~S. Park, T.~Wang, T.~I. Denk, A.~Ly, N.~Chen, Z.~Zhang, Z.~Zhang, J.~Yu, C.~Frank \emph{et~al.}, ``Noise2music: Text-conditioned music generation with diffusion models,'' \emph{arXiv preprint arXiv:2302.03917}, 2023.

\bibitem{schneider2023mo}
F.~Schneider, O.~Kamal, Z.~Jin, and B.~Sch{\"o}lkopf, ``Mo$\backslash$\^{} usai: Text-to-music generation with long-context latent diffusion,'' \emph{arXiv preprint arXiv:2301.11757}, 2023.

\bibitem{lam2023efficient}
M.~W. Lam, Q.~Tian, T.~Li, Z.~Yin, S.~Feng, M.~Tu, Y.~Ji, R.~Xia, M.~Ma, X.~Song \emph{et~al.}, ``Efficient neural music generation,'' \emph{Advances in Neural Information Processing Systems}, vol.~36, pp. 17\,450--17\,463, 2023.

\bibitem{melechovsky2023mustango}
J.~Melechovsky, Z.~Guo, D.~Ghosal, N.~Majumder, D.~Herremans, and S.~Poria, ``Mustango: Toward controllable text-to-music generation,'' \emph{arXiv preprint arXiv:2311.08355}, 2023.

\bibitem{karchkhadze2024multi}
T.~Karchkhadze, M.~R. Izadi, K.~Chen, G.~Assayag, and S.~Dubnov, ``Multi-track musicldm: Towards versatile music generation with latent diffusion model,'' \emph{arXiv preprint arXiv:2409.02845}, 2024.

\bibitem{evans2024stable}
Z.~Evans, J.~D. Parker, C.~Carr, Z.~Zukowski, J.~Taylor, and J.~Pons, ``Stable audio open,'' \emph{arXiv preprint arXiv:2407.14358}, 2024.

\bibitem{lin2024vmas}
Y.-B. Lin, Y.~Tian, L.~Yang, G.~Bertasius, and H.~Wang, ``Vmas: Video-to-music generation via semantic alignment in web music videos,'' \emph{arXiv preprint arXiv:2409.07450}, 2024.

\bibitem{muvi2024}
R.~Li, S.~Zheng, X.~Cheng, Z.~Zhang, S.~Ji, and Z.~Zhao, ``Muvi: Video-to-music generation with semantic alignment and rhythmic synchronization,'' \emph{arXiv preprint arXiv:2410.12957}, 2024.

\bibitem{jasco}
O.~Tal, A.~Ziv, I.~Gat, F.~Kreuk, and Y.~Adi, ``Joint audio and symbolic conditioning for temporally controlled text-to-music generation,'' \emph{arXiv preprint arXiv:2406.10970}, 2024.

\bibitem{Wu2023MusicControlNet}
S.-L. Wu, C.~Donahue, S.~Watanabe, and N.~J. Bryan, ``Music controlnet: Multiple time-varying controls for music generation,'' 2023.

\bibitem{Novack2024Ditto}
Z.~Novack, J.~McAuley, T.~Berg-Kirkpatrick, and N.~J. Bryan, ``{DITTO}: Diffusion inference-time t-optimization for music generation,'' in \emph{International Conference on Machine Learning (ICML)}, 2024.

\bibitem{Novack2024Ditto2}
------, ``{DITTO-2}: Distilled diffusion inference-time t-optimization for music generation,'' in \emph{International Society of Music Information Retrieval (ISMIR)}, 2024.

\bibitem{liu2024mumu}
S.~Liu, A.~S. Hussain, Q.~Wu, C.~Sun, and Y.~Shan, ``Mumu-llama: Multi-modal music understanding and generation via large language models,'' \emph{arXiv preprint arXiv:2412.06660}, 2024.

\bibitem{wang2024multimodal}
B.~Wang, L.~Zhuo, Z.~Wang, C.~Bao, W.~Chengjing, X.~Nie, J.~Dai, J.~Han, Y.~Liao, and S.~Liu, ``Multimodal music generation with explicit bridges and retrieval augmentation,'' \emph{arXiv preprint arXiv:2412.09428}, 2024.

\bibitem{gan2020foley}
C.~Gan, D.~Huang, P.~Chen, J.~B. Tenenbaum, and A.~Torralba, ``Foley music: Learning to generate music from videos,'' in \emph{Computer Vision--ECCV 2020: 16th European Conference, Glasgow, UK, August 23--28, 2020, Proceedings, Part XI 16}.\hskip 1em plus 0.5em minus 0.4em\relax Springer, 2020, pp. 758--775.

\bibitem{di2021video}
S.~Di, Z.~Jiang, S.~Liu, Z.~Wang, L.~Zhu, Z.~He, H.~Liu, and S.~Yan, ``Video background music generation with controllable music transformer,'' in \emph{Proceedings of the 29th ACM International Conference on Multimedia}, 2021, pp. 2037--2045.

\bibitem{zhuo2023video}
L.~Zhuo, Z.~Wang, B.~Wang, Y.~Liao, C.~Bao, S.~Peng, S.~Han, A.~Zhang, F.~Fang, and S.~Liu, ``Video background music generation: Dataset, method and evaluation,'' in \emph{Proceedings of the IEEE/CVF International Conference on Computer Vision}, 2023, pp. 15\,637--15\,647.

\bibitem{arnab2021vivit}
A.~Arnab, M.~Dehghani, G.~Heigold, C.~Sun, M.~Lu{\v{c}}i{\'c}, and C.~Schmid, ``Vivit: A video vision transformer,'' in \emph{Proceedings of the IEEE/CVF international conference on computer vision}, 2021, pp. 6836--6846.

\bibitem{houlsby2019parameter}
N.~Houlsby, A.~Giurgiu, S.~Jastrzebski, B.~Morrone, Q.~De~Laroussilhe, A.~Gesmundo, M.~Attariyan, and S.~Gelly, ``Parameter-efficient transfer learning for nlp,'' in \emph{International conference on machine learning}.\hskip 1em plus 0.5em minus 0.4em\relax PMLR, 2019, pp. 2790--2799.

\bibitem{pfeiffer2020adapterfusion}
J.~Pfeiffer, A.~Kamath, A.~R{\"u}ckl{\'e}, K.~Cho, and I.~Gurevych, ``Adapterfusion: Non-destructive task composition for transfer learning,'' \emph{arXiv preprint arXiv:2005.00247}, 2020.

\bibitem{pfeiffer2020adapterhub}
J.~Pfeiffer, A.~R{\"u}ckl{\'e}, C.~Poth, A.~Kamath, I.~Vuli{\'c}, S.~Ruder, K.~Cho, and I.~Gurevych, ``Adapterhub: A framework for adapting transformers,'' \emph{arXiv preprint arXiv:2007.07779}, 2020.

\bibitem{ye2023ip}
H.~Ye, J.~Zhang, S.~Liu, X.~Han, and W.~Yang, ``Ip-adapter: Text compatible image prompt adapter for text-to-image diffusion models,'' \emph{arXiv preprint arXiv:2308.06721}, 2023.

\bibitem{mou2024t2i}
C.~Mou, X.~Wang, L.~Xie, Y.~Wu, J.~Zhang, Z.~Qi, and Y.~Shan, ``T2i-adapter: Learning adapters to dig out more controllable ability for text-to-image diffusion models,'' in \emph{Proceedings of the AAAI Conference on Artificial Intelligence}, vol.~38, no.~5, 2024, pp. 4296--4304.

\bibitem{tsai2024audio}
F.-D. Tsai, S.-L. Wu, H.~Kim, B.-Y. Chen, H.-C. Cheng, and Y.-H. Yang, ``Audio prompt adapter: Unleashing music editing abilities for text-to-music with lightweight finetuning,'' \emph{arXiv preprint arXiv:2407.16564}, 2024.

\bibitem{himv2017}
\BIBentryALTinterwordspacing
S.~Hong, W.~Im, and H.~S. Yang, ``Content-based video-music retrieval using soft intra-modal structure constraint,'' 2017. [Online]. Available: \url{https://arxiv.org/abs/1704.06761}
\BIBentrySTDinterwordspacing

\bibitem{urmp2019}
\BIBentryALTinterwordspacing
B.~Li, X.~Liu, K.~Dinesh, Z.~Duan, and G.~Sharma, ``Creating a multitrack classical music performance dataset for multimodal music analysis: Challenges, insights, and applications,'' \emph{IEEE Transactions on Multimedia}, vol.~21, no.~2, p. 522–535, Feb. 2019. [Online]. Available: \url{http://dx.doi.org/10.1109/TMM.2018.2856090}
\BIBentrySTDinterwordspacing

\bibitem{tiktok2022}
\BIBentryALTinterwordspacing
Y.~Zhu, K.~Olszewski, Y.~Wu, P.~Achlioptas, M.~Chai, Y.~Yan, and S.~Tulyakov, ``Quantized gan for complex music generation from dance videos,'' 2022. [Online]. Available: \url{https://arxiv.org/abs/2204.00604}
\BIBentrySTDinterwordspacing

\bibitem{aist2021}
\BIBentryALTinterwordspacing
R.~Li, S.~Yang, D.~A. Ross, and A.~Kanazawa, ``Ai choreographer: Music conditioned 3d dance generation with aist++,'' 2021. [Online]. Available: \url{https://arxiv.org/abs/2101.08779}
\BIBentrySTDinterwordspacing

\bibitem{vtm_symbolic_2023}
L.~Zhuo, Z.~Wang, B.~Wang, Y.~Liao, C.~Bao, S.~Peng, S.~Han, A.~Zhang, F.~Fang, and S.~Liu, ``Video background music generation: Dataset, method and evaluation,'' in \emph{Proceedings of the IEEE/CVF International Conference on Computer Vision}, 2023, pp. 15\,637--15\,647.

\bibitem{Kang_2024}
\BIBentryALTinterwordspacing
J.~Kang, S.~Poria, and D.~Herremans, ``Video2music: Suitable music generation from videos using an affective multimodal transformer model,'' \emph{Expert Systems with Applications}, vol. 249, p. 123640, Sep. 2024. [Online]. Available: \url{http://dx.doi.org/10.1016/j.eswa.2024.123640}
\BIBentrySTDinterwordspacing

\bibitem{bgm909_2024}
S.~Li, Y.~Qin, M.~Zheng, X.~Jin, and Y.~Liu, ``Diff-bgm: A diffusion model for video background music generation,'' 2024.

\bibitem{nes_2024}
\BIBentryALTinterwordspacing
I.~Cardoso, R.~O.~Moraes, and L.~N.~Ferreira, ``The nes video-music database: A dataset of symbolic video game music paired with gameplay videos,'' in \emph{Proceedings of the 19th International Conference on the Foundations of Digital Games}, ser. FDG 2024.\hskip 1em plus 0.5em minus 0.4em\relax ACM, May 2024, p. 1–6. [Online]. Available: \url{http://dx.doi.org/10.1145/3649921.3650011}
\BIBentrySTDinterwordspacing

\bibitem{solovyev2023benchmarks}
R.~Solovyev, A.~Stempkovskiy, and T.~Habruseva, ``Benchmarks and leaderboards for sound demixing tasks,'' 2023.

\bibitem{giannakopoulos2015pyaudioanalysis}
T.~Giannakopoulos, ``pyaudioanalysis: An open-source python library for audio signal analysis,'' \emph{PloS one}, vol.~10, no.~12, p. e0144610, 2015.

\bibitem{kong2020panns}
Q.~Kong, Y.~Cao, T.~Iqbal, Y.~Wang, W.~Wang, and M.~D. Plumbley, ``Panns: Large-scale pretrained audio neural networks for audio pattern recognition,'' \emph{IEEE/ACM Transactions on Audio, Speech, and Language Processing}, vol.~28, pp. 2880--2894, 2020.

\bibitem{russell1980circumplex}
J.~A. Russell, ``A circumplex model of affect.'' \emph{Journal of personality and social psychology}, vol.~39, no.~6, p. 1161, 1980.

\bibitem{hung2021emopia}
H.-T. Hung, J.~Ching, S.~Doh, N.~Kim, J.~Nam, and Y.-H. Yang, ``Emopia: A multi-modal pop piano dataset for emotion recognition and emotion-based music generation,'' \emph{arXiv preprint arXiv:2108.01374}, 2021.

\bibitem{defossez2022high}
A.~D{\'e}fossez, J.~Copet, G.~Synnaeve, and Y.~Adi, ``High fidelity neural audio compression,'' \emph{arXiv preprint arXiv:2210.13438}, 2022.

\bibitem{kay2017kineticshumanactionvideo}
\BIBentryALTinterwordspacing
W.~Kay, J.~Carreira, K.~Simonyan, B.~Zhang, C.~Hillier, S.~Vijayanarasimhan, F.~Viola, T.~Green, T.~Back, P.~Natsev, M.~Suleyman, and A.~Zisserman, ``The kinetics human action video dataset,'' 2017. [Online]. Available: \url{https://arxiv.org/abs/1705.06950}
\BIBentrySTDinterwordspacing

\bibitem{hu2021lora}
E.~J. Hu, Y.~Shen, P.~Wallis, Z.~Allen-Zhu, Y.~Li, S.~Wang, L.~Wang, and W.~Chen, ``Lora: Low-rank adaptation of large language models,'' \emph{arXiv preprint arXiv:2106.09685}, 2021.

\bibitem{brown2020language}
T.~Brown, B.~Mann, N.~Ryder, M.~Subbiah, J.~D. Kaplan, P.~Dhariwal, A.~Neelakantan, P.~Shyam, G.~Sastry, A.~Askell \emph{et~al.}, ``Language models are few-shot learners,'' \emph{Advances in neural information processing systems}, vol.~33, pp. 1877--1901, 2020.

\bibitem{liu2022design}
V.~Liu and L.~B. Chilton, ``Design guidelines for prompt engineering text-to-image generative models,'' in \emph{Proceedings of the 2022 CHI conference on human factors in computing systems}, 2022, pp. 1--23.

\bibitem{white2023prompt}
J.~White, Q.~Fu, S.~Hays, M.~Sandborn, C.~Olea, H.~Gilbert, A.~Elnashar, J.~Spencer-Smith, and D.~C. Schmidt, ``A prompt pattern catalog to enhance prompt engineering with chatgpt,'' \emph{arXiv preprint arXiv:2302.11382}, 2023.

\bibitem{doh2023lp}
S.~Doh, K.~Choi, J.~Lee, and J.~Nam, ``Lp-musiccaps: Llm-based pseudo music captioning,'' \emph{arXiv preprint arXiv:2307.16372}, 2023.

\bibitem{adamw}
F.~H. Ilya~Loshchilov, ``Decoupled weight decay regularization,'' \emph{arXiv preprint arXiv:1711.05101}, 2017.

\bibitem{loshchilov2016sgdr}
I.~Loshchilov and F.~Hutter, ``Sgdr: Stochastic gradient descent with warm restarts,'' \emph{arXiv preprint arXiv:1608.03983}, 2016.

\bibitem{Novack2025Presto}
Z.~Novack, G.~Zhu, J.~Casebeer, J.~McAuley, T.~Berg-Kirkpatrick, and N.~J. Bryan, ``Presto! distilling steps and layers for accelerating music generation.'' in \emph{International Conference on Learning Representations (ICLR)}, 2025.

\bibitem{nistal2024diff}
J.~Nistal, M.~Pasini, C.~Aouameur, M.~Grachten, and S.~Lattner, ``Diff-a-riff: Musical accompaniment co-creation via latent diffusion models,'' \emph{arXiv preprint arXiv:2406.08384}, 2024.

\bibitem{wu2023large}
Y.~Wu, K.~Chen, T.~Zhang, Y.~Hui, T.~Berg-Kirkpatrick, and S.~Dubnov, ``Large-scale contrastive language-audio pretraining with feature fusion and keyword-to-caption augmentation,'' in \emph{ICASSP 2023-2023 IEEE International Conference on Acoustics, Speech and Signal Processing (ICASSP)}.\hskip 1em plus 0.5em minus 0.4em\relax IEEE, 2023, pp. 1--5.

\bibitem{kilgour2018fr}
K.~Kilgour, M.~Zuluaga, D.~Roblek, and M.~Sharifi, ``Fr$\backslash$'echet audio distance: A metric for evaluating music enhancement algorithms,'' \emph{arXiv preprint arXiv:1812.08466}, 2018.

\bibitem{ferjad2020icml}
M.~F. Naeem, S.~J. Oh, Y.~Uh, Y.~Choi, and J.~Yoo, ``Reliable fidelity and diversity metrics for generative models,'' 2020.

\bibitem{koutini2021efficient}
K.~Koutini, J.~Schl{\"u}ter, H.~Eghbal-Zadeh, and G.~Widmer, ``Efficient training of audio transformers with patchout,'' \emph{arXiv preprint arXiv:2110.05069}, 2021.

\end{thebibliography}
%

%
%
%
%

\end{document}